\documentstyle[epsfig,12pt]{article}  
\newcommand{\beq}{\begin{equation}}
\newcommand{\eeq}{\end{equation}}
\newcommand{\bea}{\vspace{0.25cm}\begin{eqnarray}}
\newcommand{\eea}{\end{eqnarray}}

\newcommand{\r}{\mbox{{\boldmath
$\rho$}}}

\newcommand{\pb}{\mbox{{\bf
p}}}

\newcommand{\kb}{\mbox{{\bf
k}}}

\def\lsim{\mathrel{\rlap{\lower4pt\hbox{\hskip1pt$\sim$}}
    \raise1pt\hbox{$<$}}}         %less than or approx. symbol
\def\gsim{\mathrel{\rlap{\lower4pt\hbox{\hskip1pt$\sim$}}
    \raise1pt\hbox{$>$}}}         %greater than or approx. symbol
\begin{document}
\vspace*{-2cm}
 
\bigskip
%%%%%%%%%%%%%%%%%%%%%%%%%%%%%%%%%%%%%%%%%%%%%%%%%%%%%%%%%%
%%%%%%%%%%%%%%%%%%%%%%%%%%%%%%%%%%%%%%%%%%%%%%%%%%%%%%%%%% 

\begin{center}

\renewcommand{\thefootnote}{\fnsymbol{footnote}}

  {\Large\bf
Anomalous mass dependence of radiative quark energy loss
in a finite-size quark-gluon plasma
% mass dependence of the gluon and photon emission
%from a fast quak
%in a finite-size quark-gluon plasma
\\
\vspace{.7cm}
  }
\renewcommand{\thefootnote}{\arabic{footnote}}
\medskip
  {\large
  P.~Aurenche$^a$ and B.G.~Zakharov$^{b}$}
  \bigskip

{\it
$^{a}$
LAPTH, Universit\'e de Savoie, CNRS,\\
BP 110, F-74941, Annecy-le-Vieux Cedex, France\\
%Laboratoire d'Annecy-le-Vieux de Physique Th\'eorique LAPTH,\\
%B.P. 110, F-74941 Annecy-le-Vieux Cedex, France\\
$^{b}$L.D. Landau Institute for Theoretical Physics,
        GSP-1, 117940,\\ Kosygina Str. 2, 117334 Moscow, Russia\\
\vspace{1.7cm}}

  {\bf Abstract}
\end{center}
{
\baselineskip=9pt
We demonstrate that for a finite-size quark-gluon plasma
the induced gluon radiation from heavy quarks is stronger 
than that for light quarks when the gluon formation length 
becomes comparable with (or exceeds) the size of the plasma. 
The effect is due to oscillations
of the light-cone wave function for the in-medium 
$q\rightarrow gq$ transition.
The dead cone model by Dokshitzer and Kharzeev 
neglecting quantum 
finite-size effects is not valid in this regime.
The finite-size effects
also enhance the photon emission from heavy quarks. 

\vspace{.5cm}
}

%-------------------------------------------------------------
\noindent{\bf 1}.
It is widely believed that the observed at RHIC strong suppression of 
high-$p_{T}$ hadrons (jet quenching) (for a review, see e.g. \cite{RHIC_data})
is due to radiative and collisional parton energy loss in 
the quark-gluon plasma (QGP) 
produced in the initial stage of $AA$-collision.
The dominating contribution to the energy loss comes from the
induced gluon emission
(for reviews, see e.g. \cite{BSZ,WK}).
The effect of collisional loss is relatively small 
\cite{Z_RAA04,Z_DE07,Z_RAA08}.
One of the interesting questions, which is important for
jet tomography of $AA$-collisions at RHIC and LHC,
is the question on the difference between gluon emission from light and
heavy quarks.
Dokshitzer and Kharzeev \cite{DK} suggested that gluon radiation 
from heavy quarks should be suppressed due to the dead cone effect.
However, the RHIC data on suppression of the non-photonic electrons 
\cite{electrons} indicate that
the energy loss 
of heavy quarks may be 
similar to that for light quarks. 
The theoretical calculations of the energy loss of one of us 
\cite{Z_DE07}
within the light-cone path integral (LCPI) approach \cite{LCPI,Z_YAF}
also are in contradiction with the predictions of \cite{DK} since
for a finite-size (FS) QGP at high 
energies the radiative loss has an anomalous mass dependence, i.e., $\Delta E_{heavy}>\Delta E_{light}$.

In the present paper we give a physical interpretation of the anomalous
mass dependence of the induced gluon emission.
We show that the effect is related to
the quantum FS effects which come into play when the gluon
formation time becomes comparable with the size of the QGP.
Physically the anomalous mass dependence 
of the induced gluon emission is due to 
oscillations of the light-cone
wave function (LCWF) for the in-medium $q\rightarrow gq$ transition.
The present paper restricts its
attention to the physical nature
of the effect. 
For this reason,
to make the analysis more transparent we,
similarly to \cite{DK},  consider a QGP with a
constant density.
The mass dependence of the jet quenching parameter
for expanding QGP will be addressed elsewhere.

\vspace{.1cm}
\noindent{\bf 2}.
We consider a fast quark with energy $E$ produced in a QGP at 
$z=0$ (we choose the $z$ 
axis along the quark momentum).
In the LCPI formalism \cite{LCPI,Z_YAF}, which we use, 
the probability of the induced gluon 
emission can be written
in terms of the Green's function
for a two-dimensional Schr\"odinger equation with the Hamiltonian
\beq
{H}=
%\frac{{\qb}^{2}}{2\mu(x)}
-\frac{1}{2\mu}
\left(\frac{\partial}{\partial \r}\right)^{2}
-\frac{i n(z)\sigma_{3}(\rho,x)}{2}+\frac{1}{L_{f}}\,,
\label{eq:10}
\eeq
where $x$ is the gluon fractional momentum, $\mu=Ex(1-x)$,
$n$ is the number medium density, $\sigma_{3}$ is the cross section
of interaction with a plasma constituent of the $q\bar{q}g$ system,
$L_{f}=2\mu/\epsilon^{2}$, $\epsilon^{2}=m_{q}^{2}x^{2}+m_{g}^{2}(1-x)$
($m_{q}$ and $m_{g}$ are the quark and gluon quasiparticle masses).
In the low density limit $L_{f}$ gives the coherence/formation length
of gluon emission in an infinite medium (the Bethe-Heitler regime).  
The three-body cross section entering
(\ref{eq:10}) can be written
as
$
\sigma_{3}(\rho,x)=\{9[\sigma_{2}(\rho)
+\sigma_{2}((1-x)\rho)]-\sigma_{2}(x\rho)\}/8\,,
$
where $\sigma_{2}(\rho)$ is the dipole cross section of
interaction with color center of the color singlet $q\bar{q}$
system. 
The Hamiltonian (\ref{eq:10}) describes in-medium evolution of the LCWF
of a fictitious $q\bar{q}g$ system.

The gluon spectrum 
for a FS medium 
can also be written in the form 
\cite{Z_Moriond98,Z_OA}
\beq
\frac{d P}{d
x}=
\int\limits_{0}^{L}\! d z\,
n(z)
\frac{d
\sigma_{eff}^{BH}(x,z)}{dx}\,,
\label{eq:20}
\eeq
\beq
\frac{d
\sigma_{eff}^{BH}(x,z)}{dx}=\mbox{Re}
\int d\r\,
\psi^{*}(\r,x)\sigma_{3}(\rho,x)\psi(\r,x,z)\,,
\label{eq:30}
\eeq
where $L$ is the quark pathlength in the medium, $\psi(\r,x)$ is 
the ordinary LCWF
for $q\rightarrow gq$ transition in vacuum,
and
$\psi(\r,x,z)$ is the
in-medium LCWF at the
longitudinal coordinate $z$ (hereafter we drop spin and color indices). 
The latter can be written as 
\bea
\psi(x,\r,z)=
\hat{U}
\left.\int\limits_{0}^{z}dz {\cal K}(\r,z|\r',0)
\right|_{\r^{'}=0}\,\,,
\label{eq:40}
\eea
where ${\cal{K}}$ is the Green's function for the Hamiltonian (\ref{eq:10}), 
and $\hat{U}$ is the spin vertex operator (its specific form can be found 
e.g. in
\cite{Z_YAF,AZ07}).
In the low density limit
 $\psi(\r,x,z)$ goes to $\psi(\r,x)$
at $z/L_{f}\gg 1$.
In this limit
(\ref{eq:30}) reduces to the ordinary 
Bethe-Heitler cross section, $d\sigma^{BH}/dx$, in the 
form derived in \cite{NPZ}.

%%%%%%%%%%%%%%%%%%%%%%%%%%%%%%%%%%%%%%%%%%%%%%%%%%%%%%%%%%%%%%
\noindent{\bf 3}.
Let us consider the qualitative pattern of gluon
emission.
% in a FS medium. 
The integral (\ref{eq:40}) is saturated at $z\gsim \bar{L}_{f}$, where
$\bar{L}_{f}$ is the effective (with the LPM effect) gluon formation 
length in an infinite medium.
The typical transverse size of the $q\bar{q}g$
system for gluon emission in 
this regime, $\bar{\rho}$, is related to $\bar{L}_{f}$
by the Schr\"odinger diffusion relation 
$\bar{\rho}^{2}\sim \bar{L}_{f}/\mu$.
In a FS medium the dynamics of the gluon emission depends crucially on
the ratio $\xi=L/\bar{L}_{f}$.
For $\xi\gg 1$
$\psi(\r,x,z)$ is very close to $\psi(\r,x,\infty)$. In this regime
the FS effects become negligible and the spectrum is close 
to that for an infinite medium
(we call this situation the infinite medium regime).
At $\xi\lsim 1$  
the effective Bethe-Heitler cross section
is chiefly controlled by the FS effects. 
In this regime 
(as in \cite{Z_OA} we call it the diffusion regime)
the dominating contribution 
comes from $N=1$ rescattering \cite{Z_OA,AZZ,OAArnold}
which gives the effective Bethe-Heitler cross section
of the order of $\xi d\sigma^{BH}/dx$.

%%%%%%%%%%%%%%%%%%%%%%%%%%%%%%%%%%%%%%%%%%%%%%%%%%%%%%%%%%%%
In the infinite medium regime the Coulomb effects are not very
important and the gluon yield can be estimated 
in the oscillator approximation (OA) which corresponds to the
parametrization $\sigma_{2}(\rho)=C_{2}\rho^{2}$. Then
the Hamiltonian takes the oscillator form with the oscillator frequency
$
\Omega=\sqrt{-i nC_{3}}/\mu
$
with
$
C_{3}=\frac{1}{8}\left\{9[1+(1-x)^{2}]-x^{2}\right\}
C_{2}\,.
$
Note that in terms of the well known BDMPS transport coefficient $\hat{q}$
\cite{BDMPS,BSZ}
$C_{2}=\hat{q}C_{F}/2nC_{A}$.
In the OA the probability of gluon emission per unit length
reads
\beq
\frac{d
P}{dxdL}=
n\frac{d\sigma^{BH}_{OA}}{dx}S(\eta)\,,
\label{eq:50}
\eeq
where 
${d\sigma^{BH}_{OA}}/{dx}=
{2\alpha_{s}C_{3}P_{gq}(x)}/{3\pi
\epsilon^{2}}\,,
$ 
is the Bethe-Heitler cross section
($P_{gq}(x)$ is the ordinary $q\rightarrow g$ splitting function), 
and $S(\eta)$ is the LPM suppression
factor given by \cite{Z_YAF} 
\beq
S(\eta)=\frac{3}{\eta\sqrt{2}}
\int\limits_{0}^{\infty}
dy
\left(\frac{1}{y^{2}}-
\frac{1}{{\rm sh}^{2}
y}\right)
\exp\left(-\frac{y}{\eta\sqrt{2}}\right)
\left[\cos\left(\frac{y}{\eta\sqrt{2}}\right)+
\sin\left(\frac{y}{\eta\sqrt{2}}\right)\right]\,
\label{eq:60}
\eeq
with
$
\eta=L_{f}|\Omega|=
\sqrt{4 n C_{3}\mu}/\epsilon^{2}\,
$.
At $\eta\gg 1$ when the LPM suppression becomes strong
from (\ref{eq:60}) one can obtain
$
S(\eta)\approx \frac{3}{\eta\sqrt{2}}
\left(1-\frac{\pi}{\eta 2\sqrt{2}}\right)
$.
Then from (\ref{eq:50}) one obtains
\beq
\frac{d P}{dx dL}
\approx
\frac{\alpha_{s}P_{gq}(x)}{2\pi}
\sqrt{\frac{2nC_{3}}
{\mu}}
\left
[1-\frac{\pi\epsilon^{2}}
{4\sqrt{\mu nC_{3}}}
\right]\,.
\label{eq:70}
\eeq
Using (\ref{eq:70}) we obtain  
the heavy-to-light mass suppression $K$-factor
\beq
K\approx 1-\frac{\pi (M_{Q}^{2}-m_{q}^{2})x^{3/2}}
{4\sqrt{E(1-x)nC_{3}}}
\,,
\label{eq:80}
\eeq
where $M_{Q}$ is the heavy quark mass.
Thus, we see that, similarly to the ordinary Bethe-Heitler cross section,
in the infinite medium regime with strong LPM suppression
the gluon yield falls with quark mass.
Note that the analysis \cite{DK} is supposed to be applicable namely
to the infinite medium regime. In our notation the mass $K$-factor
obtained in \cite{DK} reads (in \cite{DK} the light quark and gluon
are massless)
\beq
K_{DK}=
\left[
1+\frac{M_{Q}^{2}x^{3/2}}
{3\sqrt{EnC_{2}/2}}\right]^{-2}\,.
\label{eq:90}
\eeq
Both (\ref{eq:80}) and (\ref{eq:90}) are obtained for strong 
LPM suppression using the OA.
Some difference between the two expressions is not surprising
since our formula, contrary to that of \cite{DK},
is obtained with accurate treatment of the mass dependence of the oscillator
Green's function.

%%%%%%%%%%%%%%%%%%%%%%%%%%%%%%%%%%%%%%%%%%%%%%%%%%%%%%%%%%%%%%
Let us now consider the diffusion situation.
For a given $L$ the boundary $x$ for onset of this regime
can be obtained using the OA estimate $\bar{L}_{f}\sim
\mbox{min}(L_{f},|\Omega|^{-1})$ \cite{Z_OA}. 
Assuming that
the LPM effect is strong  
in the infinite medium regime (it means that $\bar{L}_{f}\sim |\Omega|^{-1}$) 
one obtains that the diffusion situation corresponds to
$x\gsim nC_{3}L^{2}/E$ (we assume that $x$ is small). 
Of course, it is only a crude estimate, and
as will be seen from our numerical results in reality 
the FS effects becomes important at much smaller $x$.

Neglecting a small contribution from $N\ge 2$ rescatterings
the effective cross section (\ref{eq:30}) for 
Debye potential with screening mass $m_{D}$ can be written in momentum space 
as
\cite{Z_OA,Z_kin}
\beq
\frac{d\sigma_{eff}^{BH} (x,z)}{dx}=
%\frac{
\frac{\alpha_{s}^{3}C_{T} P_{gq}(x)}{\pi^{2}C_{F}}
\left[F(1,z)+F(1-x,z)-F(x,z)/9\right]\,,
\label{eq:100}
\eeq
\beq
F(y,z)=\int \frac{d\kb d\pb}{(k^{\,2}+m_{D}^{2})^{2}}
H(y\kb,\pb)
\cdot\left[1-
\cos\left((p^{\,2}+\epsilon^{2})\rho_{d}^{2}(z)
\right)
\right]\,,
\label{eq:110}
\eeq
\beq
H(\kb,\pb)=
\frac{\pb^{\,2}}
{(\pb^{\,2}+\epsilon^{2})^{2}}-
\frac{(\pb-\kb)\pb }
{(\pb^{\,2}+\epsilon^{2})(
(\pb-\kb)^{2}+\epsilon^{2})
}\,,
\label{eq:120}
\eeq
where $C_{T,F}$ are the plasma constituent and quark Casimirs, 
$\rho_{d}(z)=\sqrt{{z}/{2\mu}}$ is the diffusion radius.
We represent $F$ as a sum  $F=F_{0}+\delta F$, where $F_{0}=F(\epsilon=0)$, and 
$\delta F$ is the mass correction. 
In the massless limit the momentum integration in (\ref{eq:110}) gives 
$F_{0}(y,z)=\pi^{3}y^{2}\rho_{d}^{2}(z)/2$. 
This leads to the spectrum \cite{Z_OA}
\beq
\left.\frac{dP_{N=1}}{dx}\right|_{\epsilon=0}=
\frac{\pi n L^{2}\alpha_{s}^{3} C_{T}P_{gq}(x)[1+(1-x)^{2}-x^{2}/9]}
{8C_{F}Ex(1-x)}\,.
\label{eq:130}
\eeq

An exact analytical calculation of $\delta F$ is impossible.
We have performed approximate calculation
of $\delta F$  for $\xi\gg 1$.
Keeping only the terms with large logarithms one can obtain
in this limit
\bea
\delta F(y,z)\approx
\frac{\pi^{2}\epsilon^{2}\rho_{d}^{4}(z)y^{2}}{2}
\left\{
2\log^{2}\left(\frac{1}{\epsilon^{2}\rho_{d}^{2}(z)}\right)
+\log\left(\frac{1}{\epsilon^{2}\rho_{d}^{2}(z)}\right)
\log\left(\frac{\epsilon^{2}}{y^{4}m_{D}^{4}\rho_{d}^{2}}\right)
\right.\nonumber\\
\left.
-3\log\left(\frac{1}{\epsilon^{2}\rho_{d}^{2}(z)}\right)
-\frac{y^{2}m_{D}^{2}}{\epsilon^{2}}
\log\left(\frac{1}{\epsilon^{2}\rho_{d}^{2}(z)}\right)
\right\}\,.
\label{eq:140}
\eea
Then, neglecting in (\ref{eq:140}) the linear subleading logarithms,
we obtain for the mass correction to the spectrum
\beq
\delta \frac{dP_{N=1}}{dx}\approx
\frac{\alpha_{s}^{3}
C_{T}P_{gq}(x)[1+(1-x)^{2}-x^{2}/9]Ln\epsilon^{2}\rho_{d}^{4}(L)}
{2C_{F}}
\log^{2}\left(\frac{1}{\epsilon^{2}\rho_{d}^{2}(L)}\right)\,.
\label{eq:150}
\eeq

To calculate the $N=1$ term in the OA one should replace in 
(\ref{eq:110}) 
$
{1}/{(\kb^{2}+m_{D}^{2})^{2}}
$
by
$
{2\hat{q}\delta(\kb)}/{n\alpha_{s}^{2}C_{A}C_{T}}\,.
$
In this case the $N=1$ contribution vanishes in the massless limit 
\cite{Z_OA,AZZ,OAArnold}, and all 
the contribution comes from the mass correction 
\beq
\delta \frac{dP_{N=1}^{OA}}{dx}\approx
\frac{4\hat{q}L\alpha_{s}
P_{gq}(x)[1+(1-x)^{2}-x^{2}/9]\epsilon^{2}\rho_{d}^{4}(L)}
{6\pi C_{A}C_{F}}
\log\left(\frac{1}{\epsilon^{2}\rho_{d}^{2}(L)}\right)\,.
\label{eq:160}
\eeq
Thus, one sees that in the deep diffusion regime the gluon yield
has anomalous mass dependence both for the realistic Hamiltonian and in
the OA.

In terms of the representation (\ref{eq:30}) 
the different mass dependence in the diffusion and infinite 
medium regimes is due to
qualitatively different character of the $\rho$-dependence of the
in-medium LCWF in these two situations. In the infinite medium regime
$\psi(x,\r,z)$, like $\psi(x,\r)$, is a smooth exponentially decreasing
function of $\rho$. In this case the probability of gluon emission
decreases with quark mass due to a reduction of the dominating
$\rho$ scale in (\ref{eq:30}). Naively one could expect 
that the same mechanism should suppress 
the gluon emission from heavy quarks in the diffusion case.
However, this is not true since in the diffusion
regime the in-medium LCWF entering the integrand of 
(\ref{eq:30}) becomes oscillating in $\rho$. 
It is well seen from Fig.~1 where we plot the leading order 
radial in-medium LCWF 
for different values of the dimensionless parameter $\xi=L/L_{f}$. 
One can see that the FS effects becomes small only at $L/L_{f}\gsim 10$.
The oscillations of the in-medium LCWF suppress the probability
of gluon emission. This suppression is weaker for heavy quarks
(since the ratio $L/L_{f}$ is smaller). 
If the effect of the FS oscillations overshoots the
mass suppression of the integrand in (\ref{eq:30}) which arises 
from the ordinary
LCWF, the gluon spectrum rises with quark mass. 
Our calculations above demonstrate that namely this occurs in the
deep diffusion regime.

It is not surprising that the dead cone arguments of \cite{DK}
fail in the diffusion regime since they 
are probabilistic in nature, while the oscillations of the in-medium
LCWF leading to the anomalous mass dependence is a purely quantum effect.
It is worth noting that the upper bound on the gluon momentum
up to which the derivation of the dead cone suppression is supposed to be valid
has been obtained in \cite{DK} from a crude estimate.
In reality, as will be seen from our numerical results, 
the applicability region of the 
infinite medium approximation turns out to be considerably narrower,
and in a broad kinematical range
of gluon energy instead of the dead cone
suppression we have an enhancement of the gluon yield.

Note that, similarly to the gluon emission,
the photon radiation from massive quarks is 
also enhanced in the diffusion regime.
The $N=1$ contribution
in the massless limit has been 
calculated in \cite{Z_photon} (similarly to the gluon
spectrum it is $\propto L^{2}$). The mass correction
reads
\beq
\delta \frac{dP_{N=1}^{q\rightarrow \gamma}}{dx}\approx
\frac{e_{q}^{2}\alpha_{em}\alpha_{s}^{2}
C_{T}C_{F}n L^{3}m_{q}^{2}x[1+(1-x)^{2}] }
{16E^{2}(1-x)^{2}}
\log^{2}\left(\frac{1}{\epsilon^{2}\rho_{d}^{2}(L)}\right)\,,
\label{eq:170}
\eeq
where now $\epsilon^{2}=m_{q}^{2}x^{2}$.
Due to large cross section of the charm production
the photon radiation from $c$-quark may become an important
mechanism of the photon production at LHC energies.
Since the radiated photon has large momentum ($x\sim 1$)
measurements of the photon tagged charm production
would be especially interesting.

\noindent{\bf 4}. The above analysis is very qualitative, and
is not valid in the intermediate region $L/\bar{L}_{f}\sim 1$.
For an accurate evaluation of the mass dependence
numerical calculations are necessary.
We have performed computations (with any number of rescatterings) 
using the method suggested in \cite{Z_RAA04}. It reduces 
calculation of the spectrum to solving the Schr\"odinger equation
for the Hamiltonian (\ref{eq:10}) with a smooth boundary condition.
As in \cite{Z_DE07} 
for light quark we use the quasiparticle mass $m_{q}=0.3$ and 
for gluon $m_{g}=0.4$ GeV 
obtained in \cite{LH} from the lattice data within the quasiparticle model.
For the Debye mass we take $m_{D}=\sqrt{2}m_{g}$.
For heavy quarks we take $m_{c}=1.5$ and $m_{b}=4.5$ GeV.
We have evaluated the ratio of the gluon spectra for heavy and light quarks
for plasma temperature $T=250$ MeV and $\alpha_{s}=0.4$.    
In Fig.~2 we plot the results for
$E=20$, 50 and 100 GeV, and for the plasma thicknesses $L=2,$ 4 and 6 fm.
The value $L=2$ fm approximately corresponds to the typical 
opacity of the plasma produced at RHIC energies.
The solid and dashed curves show the results with and without the FS effects.
One can see that at $E\gsim 50$ GeV the FS effects
are very important at $L\sim 2-4$ fm. 
From the point of view of the jet quenching 
the soft region $x\lsim 0.5$ is especially important. 
Fig.~2 shows that in this region
for $L\sim 2$ fm and $E\sim 100$ GeV the FS effects
increase considerably gluon emission from heavy quarks.
The FS effects become small only at $E\sim 20$ GeV for $L\gsim 4$ fm.
Thus, Fig.~2 clearly demonstrates importance of the FS effects
in the mass dependence of the gluon yield at $L\lsim 4$ fm.

For comparison with \cite{DK}
we have also performed numerical calculations in the OA when
the rescatterings can be characterized by the 
transport coefficient $\hat{q}$.
We take $\hat{q}=0.3$ GeV$^{3}$.
At this value of $\hat{q}$ the gluon spectrum
in the Bethe-Heitler limit $x\rightarrow 0$
agrees with that calculated with accurate three-body cross section. 
In Fig.~3 we present the results for two sets of the parton masses.
The solid curves correspond to the same masses as in Fig.~2, and
the dotted ones are for massless gluon and light quark as in \cite{DK}. 
The dashed curves show the Dokshitzer-Kharzeev \cite{DK} dead cone suppression 
factor.
Comparison of the dashed and dotted curves show 
that the results of the accurate calculations
differ drastically from that predicted in the dead cone model.
The results for massive and massless gluon in Fig.~3 show
that the nonzero gluon mass (the Ter-Mikaelian effect) 
reduces the anomalous mass dependence. 
The anomalous mass dependence in Fig.~3 is stronger than in Fig.~2.
It is quite natural since in the OA the leading $N=1$ 
rescattering term simply vanishes for massless partons, contrary
to the case of realistic Hamiltonian when due to the Coulomb effects 
the single scattering
contributes even in the massless limit.

From Fig.~3 one can conclude that the applicability region of the infinite
medium approximation is considerably narrower than 
it was assumed in \cite{DK}.
The authors of \cite{DK} assume that their model is valid
for the gluon energy $\omega \lsim \hat{q}L^{2}$. 
This means that for $L\sim 5$ fm 
the mass dependence can be described in terms of the dead cone suppression 
in the whole kinematical range 
of $x$ at $E\lsim 100$ GeV. Our results shown in Fig.~3 demonstrate
that this is not true. Indeed, one can see that 
the FS effects come into play very early. 
As a result for $x\lsim 0.5$, $E\sim 100$ GeV there is no 
the dead cone suppression at all.

Our results indicate that there must not be a considerable
difference between the jet quenching for $c$ and light quarks
at $E\sim 20-50$ GeV. This energy interval is interesting from the point
of view of the nuclear modification factor $R_{AA}$ 
for the non-photonic
electrons measured at RHIC \cite{electrons}. 
If charm dominates in the region $p_{T}\lsim 8$
GeV studied in \cite{electrons}, then the observed suppression of
the non-photonic electrons is not surprising. Even if at $p_{T}\sim 5-8$
 GeV about 50\% of the electrons come from bottom quark 
\cite{bc-ratio1,bc-ratio2}
it is hardly possible to speak of a serious contradiction of the theory
and experiment. Indeed, in this scenario one can obtain the electron 
suppression about 0.3-0.5 while the data \cite{electrons}
give $R_{AA}\sim 0.1-0.55$ with large error bars. 
More accurate comparison of the energy loss models
with experiment will be possible after appearance of the data on 
the open charm production that are expected soon \cite{opencharm}.

Note that numerical calculations 
within the path integral approach performed 
in \cite{ASW} also  
indicated that in some parameter range
the average energy loss grows with quark mass.
The authors interpreted this fact as
an unphysical effect related to inapplicability of the high-energy 
approximation. 
However, the situation with applicability of the LCPI formalism 
in the kinematical region dominating the anomalous mass dependence
is in fact quite good.
Indeed, the LCPI formalism \cite{LCPI} is 
derived assuming that the longitudinal momenta are large compared to the
transverse ones and the parton masses. 
Both these conditions are well satisfied in the 
diffusion kinematical region related to the anomalous mass dependence.
In this regime the situation with the validity of the LCPI approach 
is even better than in the infinite medium regime
(which gives the mass suppressed gluon yield) where the radiated 
gluons are softer.

\vspace{.2cm}
\noindent {\bf 5}. In summary, we have shown
that for a FS QGP the induced gluon radiation
from heavy quarks becomes stronger than that for light quarks 
when the gluon formation length becomes comparable with or 
exceeds the size of the plasma.
In this regime the gluon yield is dominated by the $N=1$ rescattering.
Physically the anomalous quark mass dependence is due to oscillations
of the LCWF for the in-medium $q\rightarrow gq$ transition.
The dead cone model \cite{DK}, which neglects the quantum FS effects, 
is not valid in this regime.
The anomalous mass effect becomes more
pronounced if one neglects the Coulomb effects and 
describes the parton rescatterings in terms of 
the transport coefficient. The neglect of the gluon
quasiparticle mass also enhances the anomalous mass dependence.
 
Similarly to the gluon emission, the FS effects
enhance the photon radiation from heavy quarks. 
For this reason the photon radiation from $c$-quark
may be an important mechanism of the hard photon production at LHC.

%\vspace {.2 cm}
\noindent
{\large\bf Acknowledgements}

\noindent
BGZ thanks the LAPTH for the kind hospitality
during the time when part of this work was done. 
The work of BGZ  is supported 
in part by the program SS-3472.2008.2
and 
the LEA Physique Th\'eorique de la
Mati\'ere Condes\'ee.

\newpage
%------------------------------------------------------------------
\vspace{-.5cm}
\begin{center}
{\Large \bf Figures}
\end{center}
%------------------------------------------------------------------
\begin{figure}[ht]
\vspace{-.5cm}
\begin{center}
%\vspace{-1cm}
\epsfig{file=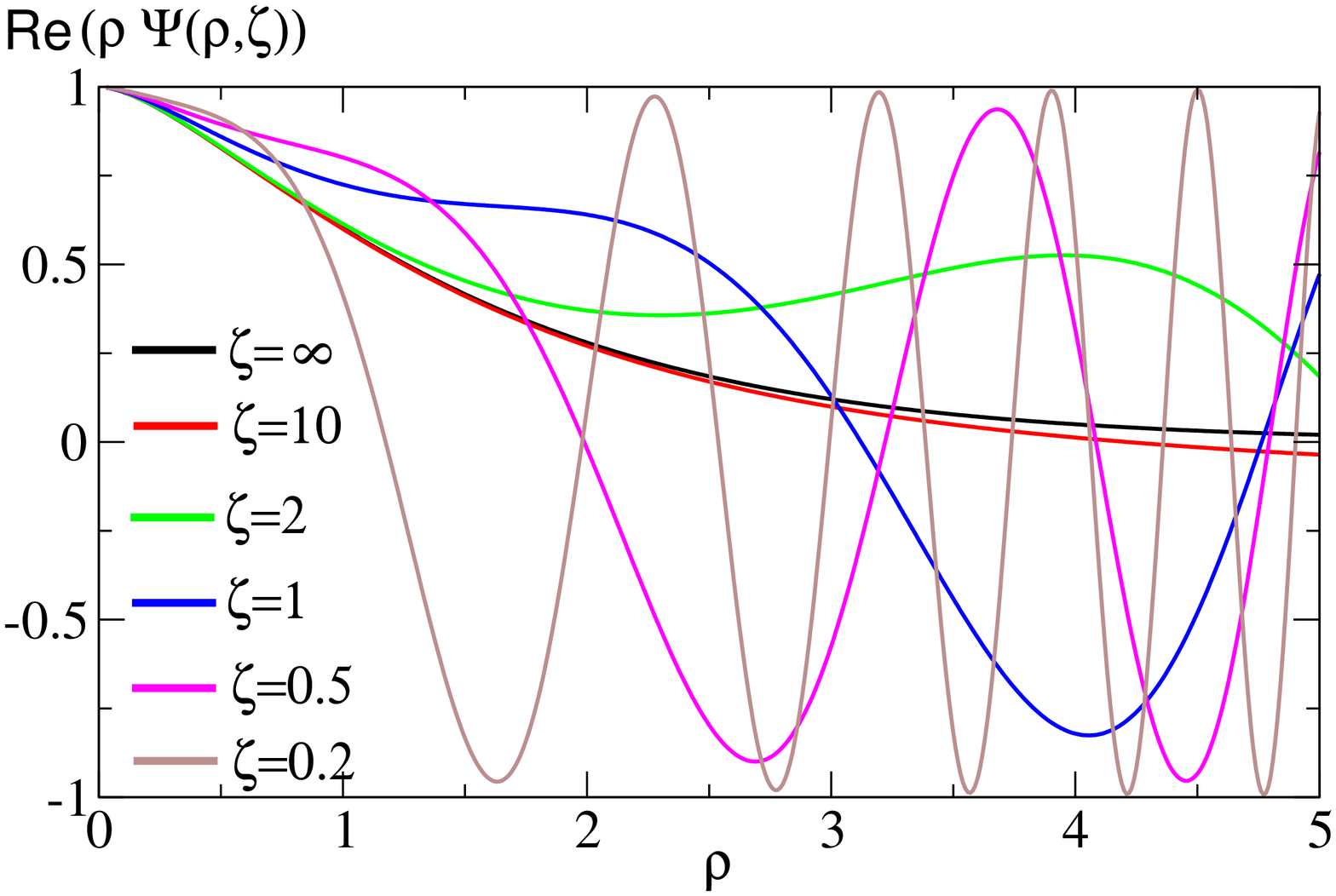,height=13cm}%,angle=-90}
\vspace{-1.5cm}
\end{center}
\caption[.]
{
%\vspace{-1.5cm}
The leading order radial LCWF for in-medium  $q\rightarrow gq$ transition
as a function of $\rho$ in units of $1/\epsilon$
for different values of the dimensionless longitudinal
coordinate $\xi=z/L_{f}$. 
}
\end{figure}
\begin{figure}[ht]
\begin{center}
%\vspace{-2cm}
\epsfig{file=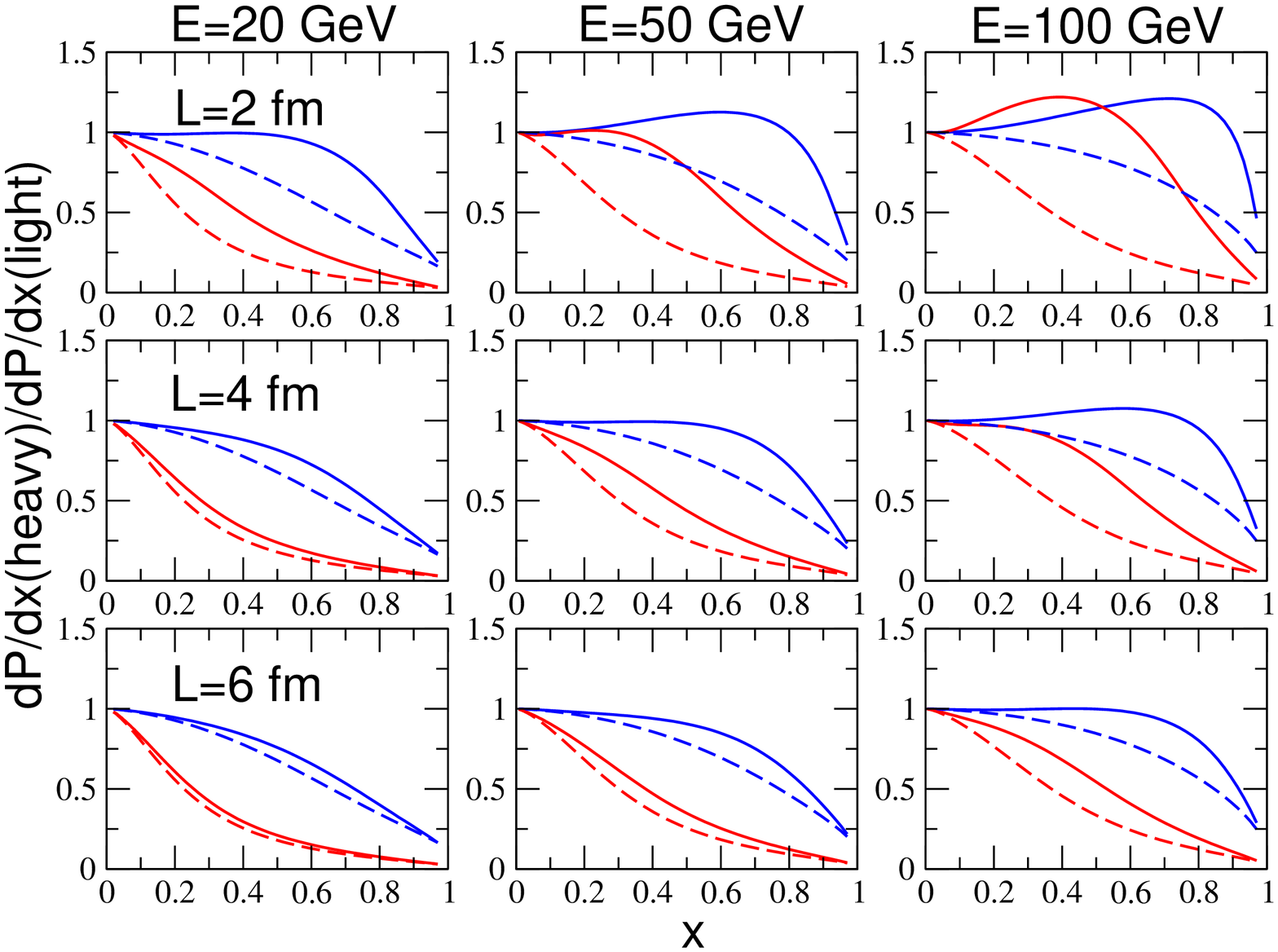,height=13cm}%,angle=-90}
\vspace{-1cm}
\end{center}
\caption[.]
{
%\vspace{-1.5cm}
The ratio of the gluon spectra for heavy and light quarks
evaluated with the dipole cross section for the Debye-screened potential.
The blue curves are for $c$-quark, the red curves are for $b$-quark.
The solid and dashed curves show the results with and without the FS effects.
All the results are for $m_{g}=0.4$ GeV.
}
\end{figure}
\newpage
\begin{figure}[ht]
\begin{center}
\epsfig{file=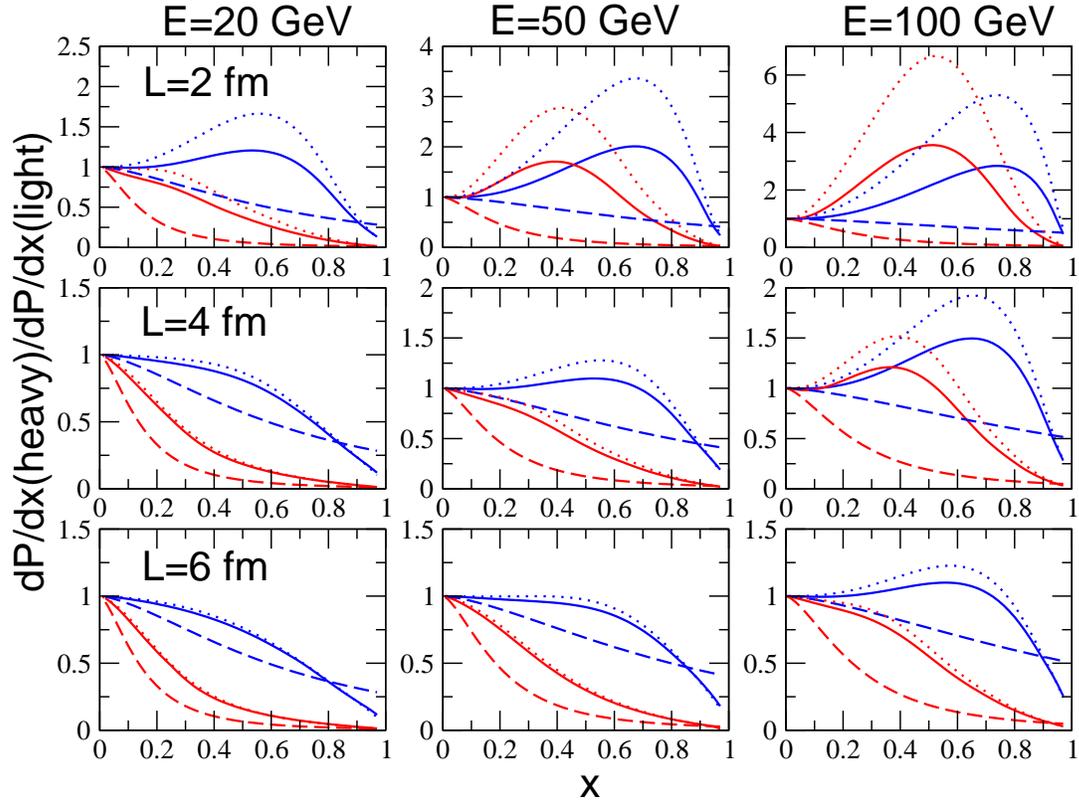,height=13cm}%,angle=-90}
\vspace{-1cm}
\end{center}
\caption[.]{
The ratio of the gluon spectra for heavy and light quarks
evaluated in the OA.
The blue curves are for $c$-quark, the red curves are for $b$-quark.
The solid curves show our results with FS effects for $m_{g}=0.4$ GeV,
and the light quark mass $m_{q}=0.3$ GeV,
and the dotted curves show our results for massless gluon
and light quark as in \cite{DK}.
The dashed curves show the Dokshitzer-Kharzeev dead cone suppression factor
(\ref{eq:90}).
}
\end{figure}

\end{document}